\RequirePackage{fix-cm}
\documentclass[twocolumn,epjc3]{svjour3} 
\RequirePackage{graphicx}
\journalname{Eur. Phys. J. C} 
\usepackage{newtxtext, newtxmath} 
\usepackage{braket,bm} 
\usepackage{multirow} 
\usepackage{enumitem} 
\usepackage[vcentermath]{youngtab} 
\usepackage{xcolor}

\begin{document}

\title{\boldmath A diquark model for the $d^\ast$(2380) dibaryon resonance?} 
\author{Avraham Gal\thanksref{e1,addr1} 
\and Marek Karliner\thanksref{e2,addr2}}
\thankstext{e1}{e-mail: avragal@savion.huji.ac.il} 
\thankstext{e2}{e-mail: marek@proton.tau.ac.il}
\institute{Racah Institute of Physics, The Hebrew University, 
Jerusalem 91904, Israel \label{addr1} \and School of Physics and Astronomy, 
Tel Aviv University, Tel Aviv 69978, Israel \label{addr2}}

\date{Received: 16 May 2019 / Accepted: 31 May 2019 / Published online: 
24 June 2019, EPJC 79:358 (2019) \\ 
https://doi.org/10.1140/epjc/s10052-019-7024-9} 

\maketitle 

\begin{abstract} 
Diquark models have been applied with varying degree of success to tetraquark 
and pentaquark states involving both light and heavy quark degrees of freedom. 
We discuss the applicability of such models to light quark dibaryons, viewed 
as three-diquark objects. Highlighting the case of the $d^\ast$(2380) 
dibaryon resonance, we demonstrate the inapplicability of diquark models 
in the light quark sector. 
\end{abstract}

\section{Introduction}
\label{intro}

The idea that diquarks ($\cal D$) play a significant role in hadron 
spectroscopy was raised by Jaffe to explain the `inverted' SU(3)-flavor 
symmetry pattern of the lowest 0$^+$ scalar-meson nonet in terms of 
tetraquarks, each made of a ${\cal D}\bar{\cal D}$ pair~\cite{Jaffe77}. 
Diquarks attracted considerable interest also in trying to understand the 
structure of the dubious $\Theta^+$(1540) pentaquark which in some experiments 
showed up as a narrow $KN$ resonant state~\cite{KL03a,KL03b,Jaffe05}. More 
recently, following the discovery of tetraquark and pentaquark structures in 
the charmed ($c$) and bottom ($b$) quark sectors, diquarks have been used in  
theoretical studies of the structure and decay patterns of such exotic states; 
for a recent review see, e.g., Refs.~\cite{ALS17,Karliner:2017qhf}. 

\begin{figure}[!ht] 
\begin{center} 
\includegraphics[width=0.48\textwidth]{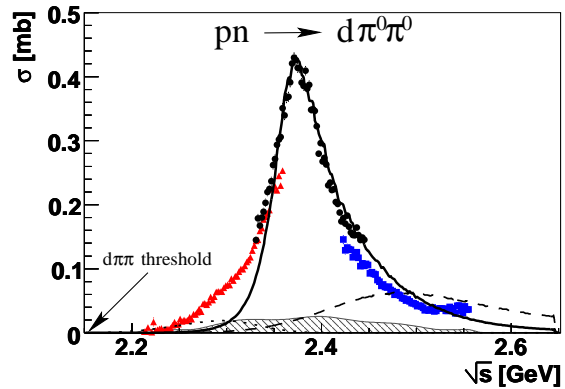} 
\caption{The $d^{\ast}$(2380) dibaryon resonance seen in the $pn\to d\pi^0
\pi^0$ reaction reported by the WASA-at-COSY Collaboration \cite{wasa11}.} 
\label{fig:WASA}
\end{center}
\end{figure}

A recent attempt to invoke diquarks to the structure of dibaryons, assuming 
that six-quark (6$q$) dibaryons consist dynamically of three diquarks, was 
made by Shi et al.~\cite{SHW19} for the $d^\ast$(2380) dibaryon resonance 
shown in Fig.~\ref{fig:WASA}. This $I(J^P)$=$0(3^+)$ fairly narrow resonance, 
peaked about 80~MeV below the $\Delta\Delta$ threshold, was observed in 
several two-pion production channels in $pn$ collisions studied by the 
WASA-at-COSY Collaboration~\cite{Clement17}. Its $I=0$ isospin assignment 
follows from balancing isospin in the $pn\to d\pi^0\pi^0$ production 
reaction, and $J^P=3^+$ spin-parity follows from the measured deuteron 
angular distribution. Subsequent measurements of $pn$ scattering and 
analyzing power~\cite{wasa14} led to a $pn$ $^3D_3$ partial-wave Argand 
diagram that supports the $d^{\ast}$(2380) dibaryon resonance interpretation. 

A major problem in understanding the structure of the $d^\ast$(2380), viewed 
as an $L=0$ $\Delta\Delta$ dibaryon, arises from its relatively small width 
$\Gamma_{d^{\ast}}$$\approx$70~MeV, see Fig.~\ref{fig:WASA}, which is by far 
smaller than twice the width of a single $\Delta$ baryon. Considering the 
reduced decay phase space available on average to a single $\Delta$ bound in 
$d^{\ast}$, its width is lowered from the free-space value of $\approx$115~MeV 
to about 80~MeV, so the problem here is how to account for a width reduction 
from about 160~MeV to $\Gamma_{d^{\ast}}$$\approx$70~MeV. This problem was 
considered in three separate approaches, the most recent of which (third one 
below) is the one we question in this note. 

(i) The $\Delta\Delta$--$\pi N\Delta$ coupled-channel hadronic calculation 
by Gal and Garcilazo~\cite{GG13,GG14} finds the $d^\ast$(2380) resonance 
at about the right position in between the corresponding thresholds, and 
with approximately the observed width. The coupled-channel nature of this 
description is essential for understanding the relatively small width in 
simple terms~\cite{Gal17}. 

(ii) Six-quark resonating-group-method calculations by Dong et 
al.~\cite{Dong16} conclude that $d^\ast$(2380) is dominated by a hidden-color 
$\Delta_8\Delta_8$ component, roughly 2:1 with respect to a `normal' 
$\Delta_1\Delta_1$ component. With color conservation forbidding the decay 
$\Delta_8 \to N_1 + \pi_1$ of a color-octet $\Delta$ to colorless hadrons, 
this leads to a substantial reduction of $\Gamma_{d^{\ast}}$, 
in good agreement with the observed value. However, the compact nature 
of the decaying $\Delta_1\Delta_1$ component introduces further reduction 
of the width, thereby resulting in over-suppression 
of $\Gamma_{d^{\ast}}$~\cite{Gal17}. 

(iii) Assuming that $d^\ast$(2380) consists of three (${\bf 6}_f, \overline{
\bf 3}_c$) flavor-color $S_{\cal D}$=1 spin diquarks, Shi et al.~\cite{SHW19} 
argued that the spatial rearrangement involved in transforming three colored 
diquarks to two color-singlet 3$q$ hadrons, with spin and flavor that identify 
them with two $\Delta$ baryons, suppresses the $\approx$160~MeV expected 
width by a factor of about 0.4. Unfortunately these authors overlooked the 
rearrangement required also in color-flavor space for a 3$\cal D$ system 
to become a $\Delta\Delta$ system. This produces another suppression factor 
of 1/9, as shown in some detail below, so the resulting width is less than 
10~MeV. 

Apart from demonstrating explicitly, based on the rough width estimate 
cited above, why a diquark model is not the right model to describe the 
$d^\ast$(2380) dibaryon resonance, the present note also discusses other 
light-quark dibaryon candidates predicted in this diquark model. It is 
concluded that diquark models in general are inappropriate for describing 
light quark dibaryons.

\section{Classification of nonstrange dibaryon candidates} 
\label{sec:L=0} 

The quark-quark ($qq$) interaction is particularly strong in the anti-triplet 
antisymmetric color state $\bar{\textbf{3}}_c$~\cite{Jaffe05}. Hence, we limit 
the discussion to $\bar{\textbf{3}}_c$ diquarks. For a nonstrange $S$-wave 
diquark, requiring antisymmetry in the combined spin-isospin-color space 
leaves one with just two spin-isospin options: $S_{\cal D},I_{\cal D}
$=0,0 scalar diquarks and $S_{\cal D},I_{\cal D}$=1,1 vector diquarks. 

Consider first a state consisting of three scalar diquark bosons, 
antisymmetrized in color space to yield a color-confined singlet 
${\textbf{1}}_c$ wavefunction. Bose-Einstein statistics then imposes 
antisymmetry on the three-diquark space wavefunction. Based on the 
experience gained in early triton binding energy calculations~\cite{DB67}, 
an antisymmetric three-body spatial wavefunction is unlikely to support 
a bound state on its own. This suggests that by trying to construct 
a dibaryon from three scalar diquarks one overlooks an important 
aspect of the dynamics. The most likely culprit is the implicit assumption 
that in order to satisfy spin-statistics one may ignore the diquarks' 
substructure and treat them all as elementary bosons. This is a rather 
dubious presumption, because in a hadron consisting of only light 
quarks there is a sole dynamical scale -- $\Lambda_{\hbox{\tiny QCD}}$. 
In the following we limit the discussion of dibaryon candidates to vector 
$\bar{\textbf{3}}_c$ diquarks.

Manipulations with $S_{\cal D},I_{\cal D}$=1,1 vector diquarks are a bit more 
involved. For a symmetric 3$\cal D$ space wavefunction, with orbital angular 
momentum $L=0$ in mind, the spin-isospin degrees of freedom have to be 
considered explicitly in forming together with a ${\textbf{1}}_c$ color 
wavefunction a totally symmetric 3$\cal D$ wavefunction. This is expressed 
schematically in terms of a product of two antisymmetric components:  
\begin{equation} 
~~~~~~~~~\left[ ~ {\yng(1,1,1)}_{~S,I} ~ \otimes ~~~ {\yng(1,1,1)}_{~c} ~ 
\right]_{~ \yng(3)} \,\, .
\label{eq:symm} 
\end{equation}     
The (1,1,1)$_{S,I}$ Young tableaux stands for the ${\textbf{84}}_{S,I}$ 
antisymmetric representation of SU(9)=SU(3)$_S \otimes$ SU(3)$_I$, where each 
of the vectors $\textbf S$ and $\textbf I$ is classified in the triplet 
$\yng(1)$ representation of the respective SU(3). This spin-isospin 
Young tableaux consists of three direct product terms: 
\begin{equation} 
{\yng(3)}_{~S}\otimes{\yng(1,1,1)}_{~I} + 
{\yng(2,1)}_{~S}\otimes{\yng(2,1)}_{~I} + 
{\yng(1,1,1)}_{~S}\otimes{\yng(3)}_{~I} 
\label{eq:prod1} 
\end{equation} 
with $S,I$ values given respectively by 
\begin{equation} 
~~~~~~1,0~~3,0~~~~+~~~~1,1~~1,2~~2,1~~2,2~~~~+~~~~0,1~~0,3~. 
\label{eq:prod2} 
\end{equation} 
Some of these 3$\cal D$ $S,I$ combinations, specifically 1,1 and 2,2, are 
spurious in terms of the underlying 6$q$ wavefunctions which are obtained 
from the following product: 
\begin{equation} 
~~~~~~~~~\left[ ~ {\yng(3,3)}_{~S,I} ~ \otimes ~~~ {\yng(2,2,2)}_{~c} ~ 
\right]_{~ \yng(1,1,1,1,1,1)} \,\, , 
\label{eq:antisym} 
\end{equation} 
where the (3,3)$_{S,I}$ Young tableaux stands for the ${\textbf{50}}_{S,I}$ 
representation of the standard SU(4)=SU(2)$_S \otimes$ SU(2)$_I$ for 
spin-1/2 and isospin-1/2 quarks. The $S,I$=3,0 dibaryon candidate in this 
6$q$ scheme was calculated to lie more than 150 MeV above the $d^\ast$(2380) 
dibaryon resonance~\cite{PPL15} which casts doubts on any attempt to ascribe 
a dominantly hexaquark structure to the observed $d^\ast$(2380).\footnote{The 
6q nonstrange dibaryons considered in Ref.~\cite{PPL15} coincide with those 
predicted long ago by Dyson and Xuong~\cite{Dyson64} who identified some of 
them with states observed near the $NN$ and $\Delta N$ thresholds, including 
the deuteron. With this remarkable insight, their predicted $S,I$=3,0 dibaryon 
came out just 30~MeV below the $d^\ast$(2380).}

\section{Dibaryon masses and rearrangement factors} 
\label{sec:disc} 

We focus now on the $I=0$ $L=0$ $S=3$ 3$\cal D$ state identified in 
Ref.~\cite{SHW19} with the $I=0$ $J^P = 3^+$ $d^\ast$(2380) dibaryon 
resonance. Its mass value was reproduced there by using an effective diquark 
mass plus color-electric and color-spin interaction matrix elements deduced 
from applying scalar and vector diquark models in the charmed sector, above 
2~~GeV. The applicability of these diquark mass and interaction parameters 
to the light-quark sector is questionable. Nevertheless based on such 
reproduction of the $d^\ast$(2380) mass, we ask where the $I=0$ 
$J^P=1^+$ deuteron-like and the $I=1$ $J^P=0^+$ virtual-like $NN$ states 
are located in this 3$\cal D$ model. Identifying these states with the 
$I=0$ $S=1$ and the $I=1$ $S=0$ states of the ${\textbf{84}}_{S,I}$ SU(9) 
representation discussed in the previous section, we evaluate their masses 
using the same $\cal D$ mass and ${\cal D}{\cal D}$ interaction parameters 
used by Shi et al.~\cite{SHW19} to evaluate the location of $d^\ast$(2380). 
Details are given here in the Appendix. The deuteron-like state $d$ is found 
then 263~MeV below the $d^\ast$(2380), about 245~MeV above the physical 
deuteron, with the virtual-like state $v$ further 53~MeV down below $d$. 
However, no resonance feature in the corresponding $I=0$ $J^P =1^+$ and 
$I=1$ $J^P =0^+$ $NN$ partial-wave phase shifts up to at least $E_{\rm cm}
$=2.4~GeV has ever been observed without any doubt~\cite{SAID19}.   

Next we evaluate the rearrangement factors involved in transforming the 
3$\cal D$ model $I=0$ $L=0$ $S=3$ state to a $\Delta\Delta$ $I=0$ $J^P =3^+$ 
$d^\ast$(2380). Since the $S=3$ Pauli spin configuration is fully stretched in 
both 3$\cal D$ and $\Delta\Delta$ bases, the spin rearrangement factor 
is simply 1. This is not the case for isospin and for color. Starting with 
isospin, we write schematically the 3$\cal D$ model couplings in the form 
\begin{equation} 
\left [ (I_1 = 1 \otimes I_2 = 1)_{I_{12}=1} \otimes (i_3 = \frac{1}{2} 
\otimes i_4 = \frac{1}{2})_{I_3=1} \right ]_{I=0}\, ,  
\label{eq:iso1} 
\end{equation} 
where the isospin structure of the $I_3=1$ third diquark is spelled out 
explicitly in terms of its quark component isospins $i_3 = i_4 = \frac{1}{2}$. 
We now recouple isospins, so that the quark isospin $i_3$ joins the diquark 
isospin $I_1=1$ to form a $\Delta$ baryon isospin $I_{13}=\frac{3}{2}$, and 
similarly the quark isospin $i_4$ joins the diquark isospin $I_2=1$ to form 
another $\Delta$ isospin $I_{24}=\frac{3}{2}$, viz. 
\begin{equation} 
\left [ (I_1 = 1 \otimes i_3 = \frac{1}{2})_{I_{13}=\frac{3}{2}} \otimes 
( I_2 = 1 \otimes i_4 = \frac{1}{2})_{I_{24}=\frac{3}{2}} \right ]_{I=0}\, . 
\label{eq:iso2} 
\end{equation} 
This recoupling is given by a unitary operator $U$ with matrix elements 
proportional to SU(2) 9j symbols~\cite{TdS63}\footnote{The proportionality 
constant=$\sqrt{(2 I_{12} {+} 1) (2 I_3 {+} 1) (2 I_{13} {+} 1) 
(2 I_{24} {+} 1)}{=}12$.}:    
\begin{eqnarray} 
 &&  U \Bigg(\begin{array}{rrr} I_1=1 & I_2=1 & I_{12}=1 \\ 
i_3=\frac{1}{2} & i_4=\frac{1}{2} & I_3=1 \\ 
I_{13}=\frac{3}{2} & I_{24}=\frac{3}{2} & I=0 
\end{array} 
\Bigg) = - \sqrt{\frac{1}{3}} \, .
\label{eq:Uiso} 
\end{eqnarray} 
Recoupling in color space is done by generalizing from SU(2)-isospin to 
SU(3)-color. The corresponding unitary operator matrix element is given 
by~\cite{SV18}: 
\begin{eqnarray}
 &&  U \Bigg(\begin{array}{ccc} \bar{\textbf{3}}_c & \bar{\textbf{3}}_c & 
 \textbf{3}_c  \\  \textbf{3}_c & \textbf{3}_c & \bar{\textbf{3}}_c  \\
\textbf{1}_c & \textbf{1}_c & \textbf{1}_c  
\end{array}
\Bigg)=\sqrt{\frac{{\rm dim}(\textbf{3}_c)}{{\rm dim}(\bar{\textbf{3}}_c) 
\times {\rm dim}(\bar{\textbf{3}}_c)}}=\sqrt{\frac{1}{3}}\, , 
\label{eq:Ucolor} 
\end{eqnarray} 
where the notation `dim' stands for the dimension (=3) of the marked SU(3)$_c$ 
representations. 

The combined recoupling coefficient in both isospin and color spaces is given 
by a product of the values noted in Eqs.~(\ref{eq:Uiso}) and (\ref{eq:Ucolor}) 
which amounts to $-1/3$. It enters quadratically in the evaluation of the 
$d^\ast$(2380) decay width to nucleons and pions via a $\Delta\Delta$ hadronic 
doorway state, hence the width suppression factor $1/9$ overlooked in 
Ref.~\cite{SHW19}. 

In a similar way, rearrangement factors for $d$ and $v$ to go into the 
corresponding $NN$ doorway states can also be evaluated, yielding somewhat 
smaller values of less than 0.1. This means that the widths involved in decays 
of such hypothetical dibaryons should be in the range of tens of MeV at most. 
Therefore, if the $d$ and $v$ 3$\cal D$ dibaryon states exist, they should 
have been already observed in $NN$ scattering experiments.

\section{Discussion and summary} 
\label{sec:sum} 

In this brief note we discussed the applicability of $\bar{\textbf{3}}_c$ 
diquark models to light-quark nonstrange dibaryons, following a suggestion 
made by Shi et al.~\cite{SHW19} that the observed $d^\ast$(2380) dibaryon 
is dominantly of a 3$\cal D$ structure. A useful test of any dibaryon model 
is provided by the extent to which it describes well the low lying hadronic 
spectrum. In this respect, we found that the 3$\cal D$ $I=0$ $J^P=1^+$ 
deuteron-like and the $I=1$ $J^P=0^+$ virtual-like states in the particular 
diquark model suggested by these authors are located some 200--250~MeV above 
the physical deuteron, where no hint of irregularities in the corresponding 
$NN$ phase-shift analyses exist. This demonstrates that diquark models are not 
physically appropriate models for binding six quarks into a dibaryon. Hadronic 
sizes that are relevant for binding together two baryons, particularly through 
pion exchange, are of order 1--2~fm and are considerably larger than the 
sub-fermi sizes expected for deeply bound 3$\cal D$ structures. 
This results in extremely small 6$q$ admixtures in the deuteron, 
see e.g. Ref.~\cite{Miller14} for a recent discussion. 

As for the $d^\ast$(2380) dibaryon specifically, which is observed through 
decay modes involving nucleons and pions that are consistent with a size of  
1--2~fm~\cite{Gal17}, we noted that if it were dominated by a 3$\cal D$ 
structure, its decay width would have been suppressed by at least an 
isospin-color recoupling factor of $1/9$ with respect to the initial $\Delta
\Delta$ hadronic estimate of 160~MeV width. We conclude that assigning 
a 3$\cal D$ structure to the $d^\ast$(2380) dibaryon is in serious 
disagreement with its total width $\Gamma_{d\ast}$$\approx$70~MeV.



\appendix
\section{color-spin matrix elements}
\label{appA}

Masses of nonstrange light-quark dibaryons in the diquark model of 
Ref.~\cite{SHW19} were given in Eq.~(8) there by 
\begin{equation} 
M_{3\mathcal{D}}=3M_{\mathcal{D}}+
\sum_{i \neq j} {\left (\alpha \,\langle{\bf\lambda}_i\cdot{\bf\lambda}_j 
\, {\bf s}_i \cdot {\bf s}_j \rangle + \beta \,\langle {\bf\lambda}_i \cdot 
{\bf\lambda}_j \rangle \right )} \, , 
\label{eq:M3D} 
\end{equation} 
where {\bf{$\lambda$}} denotes collectively the eight Gell-Mann SU(3) 
3$\times$3 matrices in color space and the sum on $i\neq j$ runs over 
all quark pairs, in the same diquark $\cal{D}$ as well as in different ones. 
For a diquark model, it is more appropriate to absorb same-diquark interaction 
terms into an effective diquark mass ${\tilde M}_{\cal D}$. For such quarks, 
$\langle{\bf\lambda}_i\cdot{\bf\lambda}_j\rangle=-\frac{8}{3}$ and 
$\langle{\bf s}_i\cdot{\bf s}_j\rangle=\frac{1}{4}$, hence 
\begin{equation} 
{\tilde M}_{\mathcal{D}}= M_{\mathcal{D}} + 2\left (-\frac{2}{3}\alpha 
-\frac{8}{3}\beta \right ) = 913.2~{\rm MeV}\, , 
\label{eq:MD} 
\end{equation} 
where the values of $M_{\mathcal{D}}=1032$~MeV, $\alpha$ and $\beta$ were 
taken from Ref.~\cite{SHW19}. Expression (\ref{eq:M3D}) is rewritten then 
in the form 
\begin{equation} 
M_{3\mathcal{D}}=3{\tilde M}_{\mathcal{D}}+\sum_{m \neq n} 
{\left (\frac{1}{4}\alpha \,\langle{\bf\lambda}_m\cdot{\bf\lambda}_n \, 
{\bf S}_m \cdot {\bf S}_n \rangle + \beta \,\langle {\bf\lambda}_m \cdot 
{\bf\lambda}_n \rangle \right )} \, ,
\label{eq:M3Drev} 
\end{equation} 
where the sum on $m\neq n$ runs on the three vector diquarks of which 
$d^\ast$, $d$ and $v$ are composed. To evaluate Eq.~(\ref{eq:M3Drev}) 
we note that by coupling $\bar{\textbf{3}}_c$ diquarks $m$ and $n$ to a 
$\textbf{3}_c$ ${\cal D}{\cal D}$ configuration, the color ${\cal D}{\cal D}$ 
interaction is determined by a single matrix element $\langle{\bf\lambda}_m
\cdot{\bf\lambda}_n\rangle=-\frac{8}{3}$, independently of the 3$\cal D$ 
dibaryon considered. Furthermore, for a spin-symmetric or antisymmetric 
3$\cal D$ wavefunction 
\begin{equation} 
{\bf S}^2=\left ( {\bf S}_1+{\bf S}_2+{\bf S}_3 \right )^2 = 
6 + 6\,\langle {\bf S}_m\cdot{\bf S}_n\rangle \, , 
\label{eq:S} 
\end{equation} 
so $\langle{\bf S}_m\cdot{\bf S}_n\rangle=1,-2/3,-1$ for $d^\ast$, $d$, $v$, 
respectively. The resulting color-spin contributions in Eq.~(\ref{eq:M3Drev}) 
are repulsive for $d^\ast$, about 160~MeV, becoming attractive for the 
other two dibaryon candidates, whereas the color-electric contributions are 
attractive, about $-$500~MeV independently of which dibaryon as long as all 
three of them are in the same 3$\cal D$ {\bf 1}$_c$ color representation. 
The mass values calculated in this 3$\cal D$ model are listed in 
Table~\ref{tab:app}. 

\begin{table}[!ht] 
\caption{${\cal D}{\cal D}$ color-spin and color interaction contributions 
to the listed total mass values $M_{3\cal{D}}$ of selected 3$\cal D$ 
dibaryons, using Eq.~(\ref{eq:M3Drev}) with ${\tilde M}_{\mathcal{D}}$ from 
Eq.~(\ref{eq:MD}). The $qq$ interaction parameters from Ref.~\cite{SHW19} are 
$\alpha=-39.5$~MeV, $\beta=32.15$~MeV. Masses are given in MeV.} 
\begin{center} 
\begin{tabular}{lccccc} 
\hline 
3$\cal{D}$ & ($I$,$J^P$) & 3${\tilde M}_{\cal{D}}$ & $V_{\rm color-spin}$ & 
$V_{\rm color}$ & ~~$M_{3\cal{D}}$ \\

\hline 
$d^\ast$ & (0,3$^+$) & 2740 & $-4\alpha$           & ~~$-16\beta$ & 2383 \\  
$d$      & (0,1$^+$) & 2740 & $+\frac{8}{3}\alpha$ & ~~$-16\beta$ & 2120 \\  
$v$      & (1,0$^+$) & 2740 & $+4\alpha$           & ~~$-16\beta$ & 2067 \\  
\hline 
\end{tabular}
\end{center} 
\label{tab:app} 
\end{table}

Taken at face value, the physical implications of Table~\ref{tab:app} are 
quite striking: the effective diquark mass in the model of Ref.~\cite{SHW19} 
is 913.2~MeV. There are three of them, with a total 3$\cal D$ mass of 
2740~MeV, so the model implies their binding energy into $d^*(2383)$ is about 
360~MeV. This is a huge binding energy for a system containing light quarks 
only. It entails a tiny radius $\sim$0.4 fm for the $3{\cal D}$ system, much 
smaller than anything known to occur in light quark systems. This raises 
further doubts regarding the physical basis of the model proposed in 
Ref.~\cite{SHW19}.

\end{document}